\documentclass[traditabstract]{aa} 
\usepackage{graphicx}
\usepackage{txfonts}
\begin{document}

\title{Optimal computation of brightness integrals parametrized on the unit sphere}

\author{Mikko Kaasalainen,\inst{1}
 Xiaoping Lu\inst{2}\and Anssi-Ville V\"anttinen\inst{1}}
\institute{Department of Mathematics,
Tampere University of Technology, PO Box 553, 33101 Tampere, Finland
\and
 Department of General Education, Macau University of Science and Technology,
Avenida Wai Long, Taipa, Macau}

\date{Received; accepted}

\titlerunning{Optimal brightness computation}
\authorrunning{M. Kaasalainen et al.}

\abstract{
We compare various approaches to find the most efficient method for the
practical computation of the lightcurves (integrated brightnesses) of 
irregularly shaped bodies such as asteroids at arbitrary viewing
and illumination geometries. For convex models, this reduces to the problem
of the numerical computation of an integral over a simply defined part of 
the unit sphere. We introduce a fast method, based on Lebedev
quadratures, which is optimal for both lightcurve 
simulation and inversion in the sense that it is the simplest and fastest
widely applicable procedure for accuracy levels corresponding 
to typical data noise. The method requires no tessellation of the
surface into a polyhedral approximation. At the accuracy level of 0.01 mag,
it is up to an order of magnitude faster than polyhedral 
sums that are usually applied to this problem, 
and even faster at higher accuracies. 
This approach can also be used in other similar
cases that can be modelled on the unit sphere. The method is easily implemented
in lightcurve inversion by a simple alteration of the standard
algorithm/software.}


\maketitle

\section{Introduction}
Many quantities in astronomy, physics, and other sciences are most
naturally represented as functions on the unit sphere $S^2$. Typical examples
in astrophysics are the celestial sphere or the surfaces of celestial
bodies. A standard operation for such functions is to compute their integrals
over the whole sphere or parts of it, usually with some weight functions, to
express total sums, averages, moments, and so on. Integrals over any shape
other than the sphere, but ones that can be mapped one-to-one 
onto $S^2$, are handled using
the surface element (Jacobian) as a weight function and are thus reduced
to exactly the same spherical problem.

Brightness integrals or lightcurves
are an important example of these surface integrals. As 
shown in \v{D}urech \& Kaasalainen (\cite{DurKaas}), virtually all lightcurves
of single asteroids in realistic observing geometries can be explained, i.e.,
fitted down to noise level, by convex surfaces approximating the
global shape of the target. Thus, for all photometric
purposes, such targets can be represented as convex models. 
Convex models require no ray-tracing as
their brightnesses can be expressed as integrals over 
a simply defined part of $S^2$.
What is more, most models are very conveniently and compactly
described by function series
(based on spherical harmonics) in lightcurve inversion (Kaasalainen et al.\
\cite{AA1,Icar1,IP}). One can thus, in principle, 
compute their total brightness (especially
that of models of real asteroids) at arbitrary viewing geometries
without the polyhedral surface representations that are the standard means of
integration in lightcurve simulation and inversion. In this paper, we  
show that function integration is faster and, in many ways,
simpler than polyhedral modelling. 
This is particularly advantageous for large-scale simulations and
inverse problems involving thousands of targets and the repeated computation
of millions of lightcurve points.

A traditional way to compute
surface integrals is to tessellate the surface, i.e., to represent it as a
polyhedral approximation, usually with triangular facets of 
similar size (triangulation). The integral
is then approximated as a sum over the facet areas multiplied by the
corresponding values of the integrand functions. The approximation
can be made arbitrarily accurate by making the triangulation mesh denser.
However,
in the $(\log N,\log\Delta)$-plane, where $\Delta$ is the approximation
error and $N$ the number of facets, i.e., of function evaluations required,
the accuracy
$\log\Delta(\log N)$ improves only with a linear slope of -1.
To reduce the error by a half, one must double the number
of computations. 
This is the least efficient way of computing the integrals,
equivalent to the elementary way of computing one-dimensional integrals
as Riemann sum approximations
by dividing the integration interval into equally large steps.

In the one-dimensional case of real-valued
functions on ${\mathbb R}$, or its Cartesian
multidimensional version ${\mathbb R}^n$, the standard way of improving the
computation is to use quadratures such as Gauss-Legendre (or
Gauss-Chebyshev etc., depending on weight functions). For large $n$, the only
computationally feasible way is to use Monte Carlo techniques. The sphere
$S^2$ has a markedly different topology from ${\mathbb R}^2$; using Gaussian
quadratures, some of the computations are wasted on integration
points clustered too tightly around the two poles, and the integration
scheme is thus
manifestly dependent on the coordinate frame. Evidently, the most natural
quadrature for $S^2$ is quite different from the Cartesian case.

Lebedev quadrature (Lebedev \& Laikov \cite{Leb})
is designed to be rather automatically
optimal for smooth and continuous integrand functions over the whole of $S^2$,
in the same sense that Gaussian quadrature is optimal for ${\mathbb R}$.
The Lebedev scheme combines the almost equidistant and rotationally invariant
distribution of points on $S^2$
with the efficiency of quadratures (a low vertical positioning
and a slope steeper than -1 for the $\log\Delta(\log N)$-curve).
However, it is not designed for integrals only over a part of $S^2$. For
these, we need to consider and test various
approaches to see which is the most useful one in practice. Our purpose is
to find the optimal approach for accuracy levels corresponding to typical
data noise.

This paper is organized as follows. In Section 2 we review the problem of
the computing of lightcurve or brightness integrals on $S^2$, and discuss its
numerical representation in the form of quadratures. In Section 3 we
present efficiency results from a various sample cases, and also
note a particular formulation for integrals on triaxial ellipsoids. 
In Section 4 we describe the use and efficiency
of the quadrature scheme in both inversion and simulations 
(e.g., Monte Carlo sampling). In the concluding Section 5 we sum up and
discuss some aspects of the quadrature approach.

\section{Brightness integrals as quadratures}

A detailed discussion of the theoretical aspects of the 
problem of computing the total brightness of a convex surface is 
given in Kaasalainen et al. (\cite{AA1,Icar1,IP}), so here we only
review the main points and notations relevant to our search of the
optimal method.

Let the illumination (subsolar) and viewing (sub-Earth) directions in
a coordinate system fixed to the target body be denoted by
$\omega_0,\omega\in S^2$. Here entities in $S^2$, defined by
two direction angles, are identified with
unit vectors in ${\mathbb R}^3$. Thus, e.g., the outward unit
normal vector $\eta\in S^2$ is given by $\eta=\eta(\vartheta,\psi)$ (with
$\vartheta$ measured from the pole); i.e.,
\begin{equation}
\eta_1=\sin\vartheta\cos\psi,\quad \eta_2=\sin\vartheta\sin\psi,\quad \eta_3=\cos\vartheta.
\end{equation}
Denoting the portion of the surface both visible and illuminated by $A_+$,
the total (integrated) brightness of the body is given by
\begin{equation}
L(\omega_0,\omega)=\int_{A_+} S(\mu,\mu_0,\alpha)\,G(\vartheta,\psi)\, d\sigma,
\label{lint}
\end{equation}
where
\begin{equation}
d\sigma=\sin\vartheta\,d\vartheta\,d\psi,
\end{equation}
and the scattering function $S(\mu,\mu_0,\alpha)$ is taken to depend
on the viewing geometry quantities
\begin{equation}
\mu=\omega\cdot\eta,\quad\mu_0=\omega_0\cdot\eta,\quad\cos\alpha=\omega_0\cdot\omega.
\end{equation}
Thus $A_+$ includes those points on the surface for which $\mu_0,\mu\ge 0$.
The curvature function $G\ge 0$ of the surface is
\begin{equation}
G(\vartheta,\psi)=\frac{J(\vartheta,\psi)}{\sin\vartheta},
\end{equation}
where $J:=\vert {\vec J}\vert$ is the norm of the Jacobian vector
\begin{equation}
{\vec J}(\vartheta,\psi)=\frac{\partial {\vec x}}{\partial\vartheta}\times
\frac{\partial {\vec x}}{\partial\psi},\label{jvec}
\end{equation}
where ${\vec x}(\vartheta,\psi)$ gives the surface as a function of the
surface normal direction.

The standard choice, adopted in Kaasalainen et al. (\cite{Icar1}),
for the numerical
computation of (\ref{lint}) is the tessellation of $S^2$ 
into $N$ almost equal-sized
facets approximating the surface as a polyhedron:
\begin{equation}
L(\omega_0,\omega)\approx \sum_{j=1}^N  S(\mu_0^{(j)},
\mu^{(j)},\alpha)\,G(\eta_j)\,\Delta\sigma_j,\label{newl}
\end{equation}
where $\eta_j$ is the surface normal of the facet $j$, and the tessellation
is done on $S^2$ with any standard scheme, with $\Delta\sigma_j$ as the
area of the facet $j$ on $S^2$. Thus the area of the facet $j$ on
the actual surface is $G(\eta_j)\,\Delta\sigma_j$, and this is solved for in
the inverse problem, leading to surface reconstruction via the Minkowski
procedure (Kaasalainen et al. \cite{AA1,Icar1}).

Another, more analytical option for evaluating (\ref{lint}) is to use
quadratures rather than the geometrically intuitive equal-facet
tessellation. The main principle in any quadrature
scheme is to choose the evaluation points of the integrand function
(and their corresponding weights)
in a cleverer way than a brute-force approximation of a Riemann sum.
This choice is based on an assumption of the form of the function.
Gaussian quadrature is based on the fact that
any one-dimensional integral (with finite itegration limits) over 
any polynomial $f(x)$ of $(2N-1)$th degree can be evaluated
exactly by choosing a set, independent of $f$ and tabulated in one step,
of merely $N$ unevenly distributed evaluation points or
abscissae $x_i$ (to be
scaled linearly according to the integration interval), 
and their weights $w_i$ so that
\begin{equation}
\int f(x)\, dx=\sum_i f(x_i) w_i
\end{equation}
(Arfken \cite{Arfken}), so the sum can be used as a good approximation
of integrals over any other well-behaved functions $f(x)$ (depending
on the number of abscissae chosen). 
Consequently, any two-dimensional integral 
\begin{equation}
\int_a^b\int_{c(v)}^{d(v)} f(u,v)\, du\, dv
\end{equation}
over a double polynomial $f(u,v)$ of degrees $L,M$ in $u,v$
can be evaluated by using $(L+1)(M+1)/4$ tabulated abscissae and their
weights. Here we
are interested in the spherical harmonics $Y_l^m(u,v)$ that can be
expressed as double polynomials in $u,v$. Since the $v$-part is expressed
by $e^{imv}$, i.e., two polynomials from the functions 
$\sin mv$ and $\cos mv$, and we usually
have $M=L$, the maximum degree of the spherical harmonics used, the
number of abscissae needed for a Gaussian quadrature over spherical
harmonics is $(L+1)^2/2$. The quadrature is usually a combination of
Gauss-Legendre and Gauss-Chebyshev to account for the trigonometric
functions with the corresponding weight function.

We now investigate the available options for writing Eq.\ (\ref{lint})
in a standard quadrature form.
In Kaasalainen et al. (\cite{AA1}), the analytical handling
of Eq.\ (\ref{lint}), leading to the uniqueness proof for the inverse problem,
required transforming the integrand into a coordinate system in
which $\omega_0,\omega$ are in the new $xy$-plane. This yields simple
integration limits, but comes at the price of a more complicated integrand:
\begin{equation}
L(\omega_0,\omega)=\int_{\alpha}^{\pi}\int_0^{\pi}P_R(\omega_0,\omega)
G(\vartheta,\psi)S(\mu,\mu_0,\alpha)\,d\sigma,\label{lnew}
\end{equation}
where
\begin{equation}
\mu=\sin\psi\sin\vartheta,\quad \mu_0=\sin(\psi-\alpha)\sin\vartheta,
\end{equation}
and the operator
$P_R(\omega_0,\omega)$ transforms a function into the new system
obtained by the frame rotation matrix ${\sf R}(\omega_0,\omega)$. Thus,
\begin{equation}
P_R(\omega_0,\omega)G(\vartheta,\psi)=G({\sf R}^{-1}(\omega_0,\omega)
\eta(\vartheta,\psi)),\label{grot}
\end{equation}
or, if $G$ is of the form $G=G(\{Y_l^m(\vartheta,\psi)\})$, where
$\{Y_l^m(\vartheta,\psi)\}$ denotes some set of spherical harmonics with
a number of various degrees and orders $l,m$,
\begin{equation}
P_R(\omega_0,\omega)G(\vartheta,\psi)=
G\left(\left\{\sum_{i=-m}^m r_i^{lm}(\omega_0,\omega)Y_l^i(\vartheta,\psi)\right\}\right),\label{yrot}
\end{equation}
where $r_i^{lm}(\omega_0,\omega)$ are the so-called D-matrices of
spherical harmonics (Kaasalainen et al. \cite{AA1}).

The integration limits of Eq.\ (\ref{lnew}) nominally allow two-dimensional
Gaussian quadrature, but the forms ({\ref{grot}) and (\ref{yrot}) both cause
significant computational overhead, as well as complications in inverse problem
applications. Using Eq.\ (\ref{grot}) calls for additional computations as the
evaluation points of $G(\eta)$ are redefined, and the transformation
(\ref{yrot}) introduces
additional multiplications for function values at the evaluation points.
Furthermore, the inverse problem solution typically requires the derivatives
of $L$ with respect to the rotation parameters of the target that define
$\omega_0,\omega$, and this leads to complicated
derivative chains in both (\ref{grot}) and (\ref{yrot}) which
increase the computational cost further. 
Note also that, due to the existence
of $S$ in the integrand, and the fact that we usually use functions
$\exp(Y_l^m)$ to guarantee positivity, the nominal number of quadrature
points discussed above is not sufficient for a series of degree $L$.

Still another way to do Gaussian quadrature over a part of $S^2$ is to
keep the integrand simple but to write the integration limits as, e.g.,
\begin{equation}
L=\int_{\psi_1(\omega_0,\omega)}^{\psi_2(\omega_0,\omega)}
\int_{\vartheta_1(\psi;\omega_0,\omega)}^{\vartheta_2(\psi;\omega_0,\omega)}
S(\mu,\mu_0,\alpha)\,G(\vartheta,\psi)\, \sin\vartheta\,d\vartheta\,d\psi,
\label{another}
\end{equation}
but now the boundary equations $\mu=0$, $\mu_0=0$ are implicit for both
$\psi_i$ and $\vartheta_i(\psi)$, so setting the Gaussian grid calls for
computationally expensive root-finding, and the parameter
derivatives are also somewhat cumbersome, as above.

Further overhead is caused in typical repeated
lightcurve computations when the function $G(\eta)$ cannot be evaluated
at a set of fixed $\eta_i$ in one step. This happens in the above integrals
due to the changing viewing geometry that alters the integration
limits, hence the sampling points of $G(\eta)$ for each computation of $L$;
for Eq.\ (\ref{lnew}) this is caused by 
$\alpha(\omega_0,\omega)$ in the integration limits. If the number of
lightcurve points for one target is $M$, that of computations required for
evaluating $S(\mu,\mu_0,\alpha)$ is I, and $K$ function evaluations are 
needed for determining $G(\eta)$ (with, e.g., spherical harmonics), then 
the computational cost of fixed-$G(\eta_i)$ evaluation grows
as $NMI$, while that of Gaussian quadrature grows as $NM(I+K)$, when
$N$ is the size of the $\eta_i$-grid. Since 
typically $K\gg I$, the $K/I$-fold overhead is considerable.

In view of the above, we thus discard the use of the above forms
for rendering (\ref{lint}) to a form suitable for 
Gaussian quadrature as the potentially
reduced number of evaluation points, as compared to tessellation, is not
nearly sufficient to compensate for the computational overhead in
the applications in practice we have in mind. In these, the
accuracy levels are around 1\% (0.01 mag), where a few hundred
facets suffice for tessellation computations in any case.
In applications where
high accuracy is needed, Gaussian quadrature over a part of $S^2$ via
Eqns.\ (\ref{lnew}) or (\ref{another})
can be useful as the efficiency of any quadrature grows very
fast with growing $N$ when the integrand is smooth and continuous over
the integration region.

Quadratures can also
be introduced by formally writing the integral (\ref{lint})
over the whole of $S^2$ by introducing a transformation zeroeing
$S(\mu,\mu_0,\alpha)$ in places:
\begin{equation}
\tilde S(\mu,\mu_0,\alpha)=\left\{\begin{array}{rl}
0, & \mu\le0\,\,{\rm or}\,\,\mu_0\le0\\
S(\mu,\mu_0,\alpha),& {\rm elsewhere,}
\end{array}\right.
\end{equation}
so that we have
\begin{equation}
L(\omega_0,\omega)=\int_{0}^{2\pi}\int_0^{\pi}
G(\vartheta,\psi)\tilde S(\mu,\mu_0,\alpha)\,d\sigma.\label{ls2}
\end{equation}
Now the integration limits are simple constants,
so this can be readily
computed with quadratures, and the evaluation points are the same for
each viewing geometry. The only hitch is that the integrand is usually
no longer smooth or not necessarily 
even continuous everywhere in the integration region
(in contrast with polynomials), so it is not
possible to give a simple estimate of the minimum size of the
quadrature grid. 

   \begin{figure}
   \centering
   \includegraphics[width=9cm]{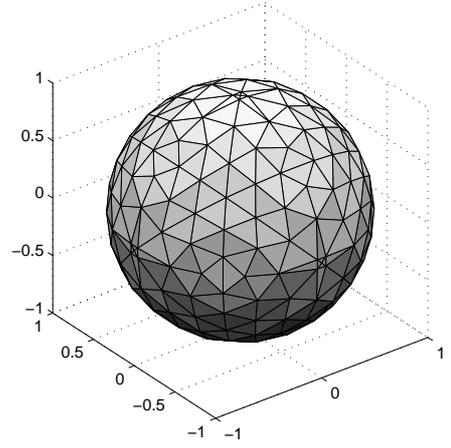}
      \caption{The distribution of Lebedev points on the unit sphere.
              }
         \label{distr}
   \end{figure}
%

While we could use Gaussian quadrature to evaluate Eq.\ (\ref{ls2}), Lebedev
quadrature is a more natural choice. It is not based on the double-integral
form, but instead evaluates integrals on the whole of $S^2$; i.e.,
\begin{equation}
I=\int_0^{2\pi}\int_0^{\pi} f(u,v)\sin u\,du\,dv\label{basicint}
\end{equation}
as an approximation
\begin{equation}
I\approx\sum_i^N f(u_i,v_i) w_i,\label{lebsum}
\end{equation}
where the evaluation points $(u_i,v_i)$ and their weights $w_i$
can be tabulated for various $N$ with standard algorithms (Lebedev \& Laikov 
\cite{Leb}). These tables can be called in the evaluations
in the same way as with other quadratures. Software for creating the
tables is available on the Internet
\footnote{see, e.g., 
{\tt 
http://people.sc.fsu.edu/$\sim$jburkardt/\\
f\_src/sphere\_lebedev\_rule/sphere\_lebedev\_rule.html}; 
{\tt
http://www.mathworks.com/matlabcentral/fileexchange/\\
27097-getlebedevsphere}
}, and sample tables and Matlab
software are available from the authors.

For Lebedev quadrature, the number of
required points in the grid 
is $2/3$ of the Gaussian case, scaling approximately 
as $(L+1)^2/3$ when the integrand $f$ is assumed to be a spherical
harmonics series with maximum degree $L$ so that the quadrature yields
the exact value of the integral
(Lebedev \& Laikov \cite{Leb}, Wang et al. \cite{LebGauss}). 
Lebedev quadrature 
is also likely to give more accurate values than a Gaussian one in the case
of a non-smooth integrand because of its almost even distribution of points on
$S^2$ that also makes
the computation virtually independent of the coordinate frame.
This is depicted in Fig.\ \ref{distr};
vertices are the Lebedev points, and the connecting edges are added by
Delaunay triangulation
to illustrate how the node distribution somewhat resembles that of typical
tessellation.

The derivation of the abscissae and weights is 
more complicated than in the Gaussian
one-dimensional case due to the wholesale operation over $S^2$ that takes
rotational symmetries into account. However, the tabulated
end result is just as easy to use as in the one-dimensional
Gaussian quadrature.
Lebedev quadrature is essentially an advanced method for arranging evaluation
points such that their distribution is approximately but not completely even, 
and consequently their weights $w_i$
are not uniform (in contrast with standard 
tessellation that aims at similar $\Delta\sigma$). 
This predictive use of symmetry gives a competitive edge over both
equal-facet tessellation and Gaussian quadrature.
Most of the points have
weights of roughly similar sizes, but the values of 
some weights are much smaller than the
average and, depending on the chosen number of points, some weight values may
even be negative. 

In fact, for Lebedev schemes with positive weight values,
Lebedev quadrature is formally equivalent to a customized
tessellation scheme, the weights being interchangeable with the 
facet areas of a polyhedron. Here $\sum_i w_i=4\pi$; by the construction of
Lebedev quadrature, the polyhedron approximates the unit sphere.
Even though no exact description of 
such an equivalence relation is needed for
the problem here, we note that the locations of the vertices corresponding 
to the areas can be determined by, e.g., 
solving the polyhedral Minkowski problem
(Kaasalainen et al. \cite{Icar1,IP}). Figure\ \ref{distr} portrays
a dual image of such a polyhedral tessellation, vertices corresponding to
facets. Lebedev points are thus not directly
the vertices of an optimal tessellation, but they indirectly define one for
certain cases.

Now, with
\begin{equation}
f(\eta)=G(\eta)\tilde S,\label{fgs}
\end{equation}
many of the $f(\eta_i)$ vanish, so
we take the quadrature sum only over the points for which $\mu_0,\mu>0$, just
as in the tessellation sum.

   \begin{figure}
   \centering
   \includegraphics[width=9cm]{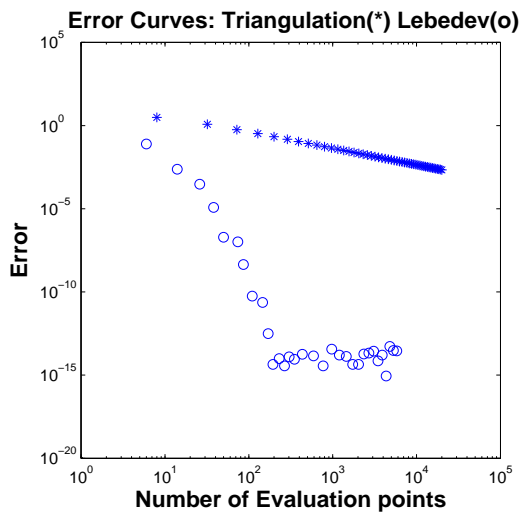}
      \caption{The error curves $\log\Delta(\log N)$ for computing
the surface integral (\ref{simple}).
Asterisks denote integration by triangulation,
while circles are for integration by quadratures.
              }
         \label{test}
   \end{figure}
%

To illustrate the remarkable power of Lebedev quadrature on
the whole of $S^2$ for a smooth and continuous integrand, we consider
the simple integral
\begin{equation}
I=\int_0^{2\pi}\int_0^{\pi}\frac{1}{2-\cos u}\sin u\,d u\, d v=2\pi\ln 3,
\label{simple}
\end{equation}
and plot the curves $\log\Delta(\log N)$ of the relative error $\Delta$
of the numerical value (from comparison with the analytical one) 
in the $(\log N,\log\Delta)$-plane
in Fig.\ \ref{test}. The error of the triangulation sum is plotted with
asterisks (with $N$ for the number of facets), 
and that of the quadrature with circles; the 
saturation of the quadrature is due to machine accuracy.
This curve also reflects the fact that quadratures over a well-behaved
integrand always eventually win over equal-facet or standard
tessellation (here we refer to this triangulation by the terms ``tessellation''
and ``triangulation'') even if there is
computational overhead, as discussed with Gaussian quadrature over a part
of $S^2$ above.

We are now left with two methods to compare: the tessellation sum
(\ref{newl}) and the Lebedev sum (\ref{lebsum}).
Our hypothesis of the superior efficiency of using $\tilde S$ with
Lebedev quadrature 
on $S^2$ rests on the assumption that the quadrature can handle the
non-smooth behaviour of $\tilde S$ without invoking too many function
evaluations; i.e., if we think in terms of an optimal tessellation via
the Lebedev procedure, it should retain its superior
properties even when shadowing is included. 
In the next section we show that this is indeed the case.
Apparently the fact that most realistic scattering models contain the
common factor $\mu\mu_0$, thus making the scattering function $S(\mu,\mu_0)$
vanish at the integration boundary, also helps to retain the accuracy in
the quadrature since the integrand is then at least continuous
on all of $S^2$.

\section{Benchmark examples}

In this section, we present some benchmark tests of the accuracy of the
methods of computing surface integrals as the size of the evaluation
grids grows.
In all figures, we have denoted the nominal number ($N$) 
of function evaluations
in the tessellation and quadrature sums. 
In brightness examples, more than half of the evaluation points
are skipped in the
computation due to $\mu\le 0$ or $\mu_0\le 0$ (i.e., $\tilde S=0$),
but the relative portions of the skipped terms are the same for
both sums, so the relative difference between the numbers of computations
stays the same.

   \begin{figure}
   \centering
   \includegraphics[width=9cm]{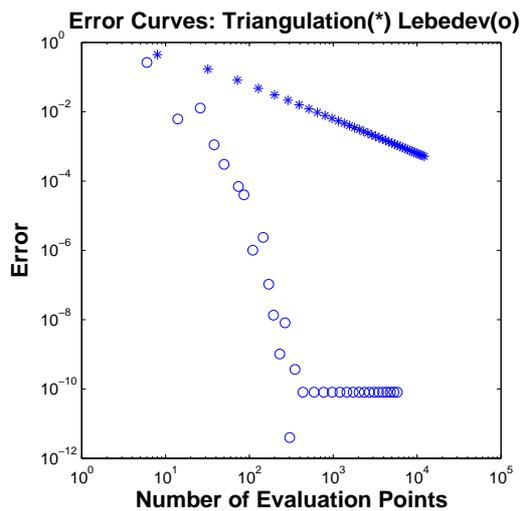}
      \caption{The error curves $\log\Delta(\log N)$ for computing
the area of an ellipsoid. Asterisks denote integration by triangulation,
while circles are for integration by quadratures.
              }
         \label{area}
   \end{figure}
%

As another example of Lebedev quadrature with smooth
integrands over the whole of $S^2$, we plot in Fig.\ \ref{area}
the relative approximation error $\Delta$
of the area $A$ of an ellipsoid against the size $N$
of the evaluation grid (asterisks for tessellation facets and
circles for Lebedev points;
$A$ is not analytically computable, but
we use a numerical elliptic integral routine for establishing the accuracy). 
We thus compute
\begin{equation}
A=\int_0^{2\pi}\int_0^{\pi} G(\vartheta,\psi)\sin\vartheta\,d\vartheta\,d\psi,
\end{equation}
where $G(\vartheta,\psi)$ for an ellipsoid with semiaxes $a,b,c$
is given in Kaasalainen et al. (\cite{AA2}):
\begin{equation}
G(\vartheta,\psi)=\left(\frac{abc}{(a\sin\vartheta\cos\psi)^2
+(b\sin\vartheta\sin\psi)^2+(c\cos\vartheta)^2}\right)^2.\label{ellg}
\end{equation}
Again, the quadrature is clearly superior to tessellation, as
expected. The saturation at high accuracy is mostly due to numerical
roundoff effects.
   \begin{figure}
   \centering
   \includegraphics[width=9cm]{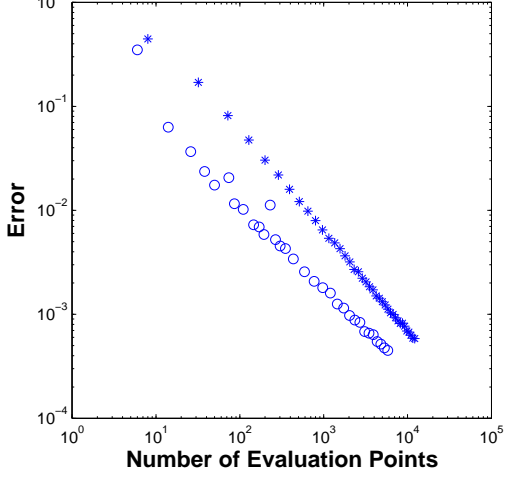}
      \caption{Average $\log\Delta(\log N)$ for geometrically scattering
ellipsoids at various random observing geometries (asterisks for tessellation,
circles for quadrature). $N$ is the nominal number of evaluation points in
a sum (over half of these are skipped).
              }
         \label{geometric}
   \end{figure}
%

The situation changes visibly when we
consider the brightness of a geometrically
scattering ellipsoid, i.e., $S=\mu$. This can be computed analytically
(Ostro \& Connelly \cite{OC}) for comparison:
\begin{equation}
L=\frac{\pi}{2}abc \Big[\sqrt{\omega^T{\sf M}\omega}+
\frac{\omega^T{\sf M}\omega_0}{\sqrt{\omega_0^T{\sf M}\omega_0}}\Big],
\end{equation}
where
\begin{equation}
{\sf M}=\left(\begin{array}{rrr}1/a^2&0&0\\
0&1/b^2&0\\
0&0&1/c^2\end{array}\right).
\end{equation}
We chose random realistic observing geometries with roughly the same
proportions of terms skipped in the quadrature (facets omitted in the
tessellation sum).
Now the integrand function $\tilde S$ is not even continuous over $S^2$,
and this shows in the quadrature error curve in Fig.\ \ref{geometric} that
is similar to the tessellation curve in its slope. However, it lies clearly
below the latter, so even now its performance is noticeably better
for our purposes.
By changing the viewpoint 90 degrees and looking at the $\Delta(N)$-curves as
plots of the inverse function $N(\Delta)$ of the
number of evaluations required for
a given accuracy (especially around the relative error 0.01), one
can clearly see the difference in computational cost 
between the two approaches.

Realistic scattering functions make $\tilde S$ continuous, which
turns out to be of an important value.
As a more general test, we examine the accuracy performance of Lebedev
and tessellation grids by computing the brightness $L$ at 
a number of random observing geometries for
general convex shapes (asteroid models obtained from real lightcurve
data). Now $G(\vartheta,\psi)$ is given by exponential spherical harmonics
series of shape models 
as used in inversion (Kaasalainen et al. \cite{Icar1}).
The scattering model used here was of the form combining the 
Lommel-Seeliger and Lambert laws,
\begin{equation}
S(\mu_0,\mu,\alpha)=f(\alpha)\mu_0\mu\left(\frac{1}{\mu+\mu_0}+c\right),
\end{equation}
which is typically used in lightcurve inversion; i.e., $f(\alpha)$ has no
influence on the plot in Fig.\ \ref{general}. The reference value of $L$ for
computing $\Delta$ was obtained by applying a dense tessellation grid.
Even though the performance
of the quadrature is not as good as for actual integrals over the whole
of $S^2$, the Lebedev-based scheme is clearly more efficient than
tessellation also in that its error curve has a steeper slope. The same
behaviour applies to, e.g., ellipsoids. Also, 
in Fig.\ \ref{general}, there are no scattered points such as the few
seen in the Lebedev curve in Fig.\ \ref{geometric}. 
This is another beneficial effect of the
continuity provided by the product $\mu\mu_0$: when 
Lebedev points are sparse, a discontinuous integrand is more 
vulnerable to the geometry of the integration depending on whether 
individual Lebedev points (and their weights)
are included in the quadrature sum or not. In general, predicting
the performance of the quadrature-based approximation 
for discontinuous integrands, especially with a low number of evaluation
points, is usually possible only by numerical tests.

   \begin{figure}
   \centering
   \includegraphics[width=9cm]{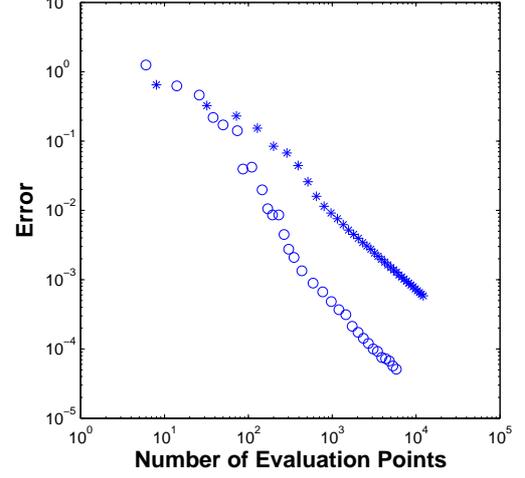}
      \caption{Average $\log\Delta(\log N)$ for general asteroid shapes
and Lommel-Seeliger/Lambert scattering at various random observing geometries.
$N$ is the nominal number of evaluation points in
a sum (over half of these are skipped).}
         \label{general}
   \end{figure}
%

The choice of the surface parametrization
determines how the quadrature evaluation points are distributed, and for
a priori known convex shapes we can, of course, use any representations
of the surface points ${\vec x}(u,v)$, not necessarily just the Gaussian image
${\vec x}(\eta)$. In such cases we write
\begin{equation}
L(\omega_0,\omega)=\int_{A_+} S(\mu_0,\mu,\alpha)\,J(u,v)
\, du\,dv,\label{ellparint}
\end{equation}
and the normal vector required for determining $\mu,\mu_0$ is
$\eta={\vec J}/J$. For example, ellipsoidal surfaces can be given in the
standard parametrization
\begin{equation}
x_1=a\sin u\cos v,\quad x_2=b\sin u\sin v,\quad x_3=c\cos u.
\end{equation}
Thus, for an ellipsoid, we have, using $u,v$ instead of $\vartheta,\psi$ in
(\ref{jvec}),
\begin{equation}
\frac{J(u,v)}{\sin u}=\sqrt{(bc\sin u\cos v)^2+(ac\sin u\sin v)^2+
(ab\cos u)^2}.\label{ellj}
\end{equation}
Since $0\le u\le\pi$ and $0\le v<2\pi$, the integral (\ref{ellparint}) is
reduced to the standard Lebedev quadrature form with an integrand
$\tilde S J(u,v)/\sin u$. As it happens, this parametrization is marginally
but consistently more efficient than the Gaussian image, so combined with
Lebedev quadrature, this is apparently the nominally 
fastest way to compute brightness integrals over ellipsoids. In practice,
we found that the choice of surface parametrization has very little effect.

\section{Efficiency in lightcurve simulation and inversion}

We have shown above that, at an accuracy level of about 1\% (corresponding
to 0.01 mag in brightness), our scheme based
on Lebedev quadrature is some five to ten times faster than triangulation, and
orders of magnitude faster at higher accuracies. Thus, for direct brightness
computations, i.e., lightcurve simulations, its efficiency is obvious.
An important point is that functions are precomputed for chosen fixed 
points on $S^2$ just as with tessellation, so the procedure introduces no
computational overhead. It is also easy to implement with Eqns.\ (\ref{lebsum})
and (\ref{fgs}) using the known 
$G(\eta)$ or $J(u,v)$ of the shape and any chosen scattering model $S$.
The increase in speed is particularly important for the
massive lightcurve simulations typically needed for estimating the performance
of large-scale sky surveys such as LSST, Pan-STARRS or Gaia. The simulations
can be carried out with shapes (i.e., $G(\eta)$) known for real asteroids
from lightcurve inversion (\v{D}urech et al.\ \cite{DAMIT}). 
Simulations with ellipsoids are fastest to run using Eqns.\
(\ref{ellj}) or (\ref{ellg}).

In lightcurve inversion, most of the computation time is spent on
evaluating lightcurves and the corresponding 
partial derivatives $\partial L/\partial P_i$
for the shape and rotation parameters $P_i$, so the enhanced
speed in brightness computation yields an equally faster inversion procedure.
The partial derivatives consume
exactly the same computation time per evaluation point $\eta_i$ 
for both quadrature and tessellation. From Eq.\
(\ref{lebsum}) we have
\begin{equation}
\frac{\partial L}{\partial P_R}=\sum_iw_iG(\eta_i)\frac
{\partial S(\mu,\mu_0,\alpha)}{\partial P_S}
\end{equation}
for rotation parameters $P_R$, and
\begin{equation}
\frac{\partial L}{\partial P_S}=\sum_iw_iS(\mu,\mu_0,\alpha)
\frac{\partial G(\eta_i)}{\partial P_R}
\end{equation}
for shape parameters $P_S$; the derivatives are
of the same form as in the tessellation
case (\ref{newl}). As before, functions (such as spherical harmonics)
used in computing $G(\eta)$ are precomputed for chosen fixed points 
on $S^2$ just as with tessellation. Lebedev quadrature
is thus easy to substitute for tessellation in the inversion procedure,
and it introduces no computational overhead, so the difference in the
computational cost between the two approaches can be read directly
from the $\Delta(N)$ (or $N(\Delta)$)-curves in Fig.\ \ref{general}. 
We plan to include a
Lebedev-based inversion version of the {\tt convexinv} 
procedure downloadable in the software section
of the DAMIT website (\v{D}urech et al. \cite{DAMIT}).

The simplest 
way to utilize the superior speed of the quadrature approach in 
lightcurve inversion is the ``black box'' mode
where the optimized $G(\eta)$ is used as an auxiliary unseen quantity, and the
output parameters are those of rotation (pole direction and period).
This is very useful in applications such as
i) searching for the correct period in sparse
photometric datasets from sky surveys
when no period is evident a priori and a wide period range must
be combed through (Kaasalainen \cite{sparseK}, \v{D}urech
et al.\ \cite{sparse1,sparse2}); ii) 
Monte-Carlo sampling to estimate the
goodness-of-fit levels of the rotation parameters and thus
their likelihood distributions; or iii)
searching for and then fixing the pole and period prior to producing a 
high-resolution version of the shape as is usually done in the
convex inversion procedure (Kaasalainen et al. \cite{Icar1}). 
Upon computing sample lightcurve inversion cases for targets previously
analysed with tessellation, and using a reduced number of Lebedev evaluation
points in accordance with Fig.\ \ref{general}, we found that the shape
results were virtually
indistinguishable from the previous ones, and the spin state
parameters were essentially the same (e.g., typically within one or two degrees
of the previous pole solution), i.e., well within uncertainty limits and
amounting to a slightly different initial guess. We emphasize
that the quadrature approach has very simple computational
plug-in properties as it has exactly the same form as, and is thus
directly interchangable with, the tessellation 
part in the software, producing essentially the same results faster.

Once the inversion procedure has provided the curvature function $G$ (in, e.g.,
the form of the coefficients of a spherical harmonics series), this
can be discretized into facet areas and normals
usable by the Minkowski procedure by 
tessellation at the desired level of resolution, and this renders the 
result as a standard polyhedral model (Kaasalainen et al.\ \cite{AA1,Icar1}).
Another way to use the Lebedev-based approach is to solve for
the separate values $G(\eta_i)$ at Lebedev points $\eta_i$ directly, in
the same way that individual facet areas can be solved for in the
tessellation-based inversion (using the exponential form 
$G(\eta_i)=\exp[a_i]$ to guarantee
positivity). Once these values are obtained, the facet areas and normals
required by the Minkowski procedure can be defined by $G(\eta_i)w_i$
(when all $w_i$ are positive) or by, e.g., using the Delaunay
triangulation of the unit sphere shown in Fig.\ \ref{distr}. The normals
of the triangles give the facet normals of the final polyhedral model, 
and its facet areas are given by $\bar G\Delta\sigma$, where $\bar G$ is
the average value of $G$ at the three vertices of the corresponding 
Delaunay triangle, and $\Delta\sigma$ is the area of the triangle. Thus
the quadrature approach provides exactly the same 
low/high-resolution options as the
standard tessellation-based procedure (e.g., {\tt convexinv)} of Kaasalainen et
al.\ (\cite{AA1,Icar1}).

As described in Kaasalainen et al. (\cite{Icar1}),
the convex-shape
procedure requires a regularization function (though usually
with a very modest weight) to make sure that the resulting shape really
is convex. This regularization is based on the three equations fulfilled
by convex surfaces:
\begin{equation}
\int_0^{2\pi}\int_0^{\pi} G(\vartheta,\psi)\eta_i(\theta,\psi)\sin\vartheta\,d\vartheta \,d\psi=0,\quad i=1,2,3;
\end{equation}
i.e., a proper curvature function $G$ has no first-degree terms $Y_1^m$ in
its representation as a spherical harmonics series. 
In regularization, the left-hand sides are enforced to be as close to zero
as possible. The regularization integral is directly in the Lebedev form,
so it is conveniently computed with the same quadrature as the lightcurve
integral. 

\section{Conclusions and discussion}

This paper complements the study of the numerical computation of lightcurve
integrals introduced and discussed 
in Kaasalainen et al.\ (\cite{AA1,Icar1}).
We have systematically investigated and compared different methods of
computing lightcurve-type integrals over varying integration regions on
the unit sphere $S^2$. For smooth and continuous integrals over the whole
of $S^2$, Lebedev quadrature is usually the optimal method, but integrals
over a part of $S^2$ depend on the case. For these integrals, the usual 
choice of (Gaussian) quadrature is computationally expensive with lightcurves 
because then i) either the 
integration limits are complicated or
they can be transformed into simpler ones only at the price of
complicating the computation of the integrand; and ii) when lightcurve 
computations are repeated for an object, functions have to be constantly
re-evaluated, whereas for tessellation and for quadrature over the
whole of $S^2$ they can be evaluated in one step. 
For lightcurve integrals, and other integrals of
similar nature, a modification of Lebedev
quadrature by zeroeing a part of the quadrature sum outperforms both 
tessellation and quadrature over a part of $S^2$
(the latter when the required relative accuracy is not 
extraordinarily high -- all realistic applications fall into this category).
The scattering function usually vanishes at the integration
boundary, which increases the efficiency of the computation as it makes the
nominal integrand continuous.

Our scheme based on Lebedev quadrature is thus the optimal method for
computating lightcurve integrals in practice. 
As a rule, Lebedev quadrature is five to ten
times faster than triangulation at the accuracy level of 1\% (0.01 mag),
and it is orders of magnitude faster at accuracies
of 0.1 \% and beyond (the slope of the
accuracy curve is close to -2 rather than -1). The superior speed of
the Lebedev approach is advantageous in any applications in lightcurve
inversion and simulation, and
it is particularly suited to the automatic {\it en masse}
analysis of thousands of targets from large-scale
surveys, such as Pan-STARRS or LSST, or for simulations requiring the
computation of a large number of lightcurve points.

Our method is very easy to substitute for tessellation in  
lightcurve simulation and inversion software since it is exactly similar
to tessellation in its computational form: it merely uses fewer
function evaluation points. 
The only thing that changes is that Lebedev
weights (constants) are substituted for the fixed areas of facets on 
the unit sphere used in tessellation-based computations, and Lebedev
points on the sphere are substituted for the fixed directions of the
facet unit normals.

\begin{acknowledgements}
This work was supported by the Academy of Finland project
``Modelling and applications of stochastic and regular surfaces in
inverse problems''. The work of X. Lu is partially supported by
grant No. 019/2010/A2 from the Science and Technology Development Fund,
MSAR. We thank an anonymous referee for valuable comments and
suggestions.
\end{acknowledgements}

\end{document}